# Hidden Degradation Costs in Energy-Cost-Only HEMS Optimisation: Study on Battery and PV Sensitivity

Dawood A Butt[1], Nandor Verba[*1]
[1]WMG, The University of Warwick, Coventry, United Kingdom
*Corresponding author: nandor.verba@warwick.ac.uk

**ABSTRACT**

Residential battery energy storage systems (BESS) are increasingly deployed alongside photovoltaic (PV) generation to reduce household energy costs under volatile time-of-use (TOU) tariffs. Model predictive control (MPC) is a widely adopted optimisation strategy for home energy management systems (HEMS), typically formulated to minimise net energy cost, subject to physical and operational constraints. However, battery degradation is rarely embedded in the optimisation objective, meaning its cost is unquantified and aggressive; high-cycle-count strategies could incur significant losses once deployed to physical systems. This paper presents a receding-horizon mixed-integer linear programming (MILP) baseline for a UK residential HEMS, using demand data from the REFIT dataset. A 3 by 3 sensitivity study is conducted across three battery sizes and three PV array sizes, with post-hoc degradation cost estimated using the Naumann stress model and rainflow cycle counting. Results show that degradation remains constant for each battery size and can exceed energy cost savings by up to 1,060 %. These results demonstrate that energy-cost-only optimisation systematically underestimates the true system cost, motivating a degradation-aware control formulation.

## 1. INTRODUCTION AND RELATED WORK

UK household energy costs have risen significantly in recent years, driven by highly volatile wholesale electricity prices. Tariffs like Octopus Agile illustrate this, launching with an initial cap of 35p/kWh, subsequently raised to 100p/kWh in October 2022 [1]. As such, intelligent residential energy control strategies capable of reducing cost and grid dependence are attracting growing research interest. As PV and BESS installation costs are decreasing, adoption of these technologies is increasing. However, where battery degradation is considered, it is typically modelled as a simple cycle limit or a capacity constraint rather than a physically grounded economic cost embedded in the optimisation objective. This paper quantifies the hidden degradation cost imposed by energy-cost-only MILP optimisation across a 3-by-3 battery and PV sizing sensitivity study, using UK residential demand data and TOU tariffs, demonstrating that naïve optimisation objectives underestimate the true system cost.

MPC is widely used in HEMS research due to its ability to coordinate PV, BESS and flexible loads under explicit constraints [2]. The optimisation horizon is typically 24 hours (96 × 15-minute steps) with a rolling re-optimisation. Recent work has started to incorporate battery degradation, though this is typically a simple constraint, such as a daily cycle limit [3] rather than a physically grounded cost. Battery degradation modelling approaches in the literature include physics-based, empirical, and data-driven methods. For this study, semi-empirical models are adopted due to compatibility with available household operational data, which lacks cell-level measurements by physics-based approaches. The DoD (depth of discharge) dependent cubic stress model [4] is validated on over two years of PV-BESS data. This is operationalised into a per-cycle degradation cost framework as a function of DoD and average SOC (state of charge) [5]. Finally, rainflow cycle counting [6] is used to extract cycles from the SOC profile for cost estimation. While recent work has begun embedding cycle constraints to limit degradation [3], the magnitude of hidden degradation cost under unconstrained energy-cost-only optimisation has not been quantified in a UK residential context with real TOU tariff data.

## 2. METHODOLOGY

The HEMS system is equipped with PV generation, BESS, and grid connection, and the MILP is formulated using CVXPY and solved with Gurobi 13.0 [7]. Household demand is sourced from the REFIT dataset [8], a UK residential load dataset which contains smart meter measurements from 20 households in Loughborough collect at 8-second resolution between 2013 and 2015. House 5 is used for this study. PV data is historical PVGIS data [9], and the import and export tariffs are Octopus Agile Region C tariff data [1], [10], and coincide with the location of the house. Household demand and PVGIS irradiance data were shifted forward six years to align with the Octopus Agile Outgoing export tariff, which started in April 2019, yielding a study period of September 2019 to July 2021. 29 February 2020 was excluded as no corresponding REFIT observation exists under the six-year offset. All data was resampled to 15-minute resolution, equating to 62,184 timesteps in total. Missing demand values (5,767 timesteps) were forward-filled, consistent with prior use of this dataset.

The objective function minimises net energy cost (1) over the study period, where $P_t^{imp}$ and -$P_t^{exp}$ are the grid import and export power (kW), $\lambda_t^{imp}$ and $\lambda_t^{exp}$ are the corresponding Agile tariff prices (£/kWh), and $\Delta t$ is the timestep duration (0.25 hours).

$$min \sum_t (P_t^{imp}\, \lambda_t^{imp} - P_t^{exp} \lambda_t^{exp}) \times \Delta t \tag{1}$$

The power balance constraint (2) ensures demand is met at every timestep, where $P_t^{load}$, $P_t^{dis}$, $P_t^{ch}$, and $P_t^{PV}$ denote household demand, battery charge, discharge, and PV generation (kW) respectively. The BESS may be charged from PV generation or grid import, and the MILP selects the optimal combination at each timestep.

$$P_t^{load} = P_t^{imp} - P_t^{exp} + P_t^{dis} - P_t^{ch} + P_t^{PV} \tag{2}$$

The SOC dynamics are defined in (3), where $SOC_t \, \epsilon[0.20, 0.80]$, $\eta_{ch} = \eta_{dis} = 0.95$ [11], and $E_{max}$ is the usable battery capacity (kWh).

$$SOC_{t+1} = SOC_t + \frac{\eta_{ch} P_t^{ch} \Delta t}{E_{max}} - \frac{P_t^{dis} \Delta t}{\eta_{dis} E_{max}} \tag{3}$$

Charge and discharge power are bounded by the inverter limit $P_{max}$ in (4).

$$0 \leq P_t^{ch} \leq P_{max};\ 0 \leq P_t^{dis} \leq P_{max} \tag{4}$$

Grid constraints (5) and (6) enforce mutual exclusivity of import and export via a binary variable $z_t \, \epsilon \, \{0, 1\}$ and $P_{grid,max} = 18.4kW$ (80A residential fuse) and $P_{exp,max} = 3.68kW$ (G98 single-phase).

$$P_t^{imp} \leq P_{grid,max} \times z_t \tag{5}$$

$$P_t^{exp} \leq P_{exp,max} \times (1 - z_t) \tag{6}$$

PV power is calculated in (7), where $G_t$ is solar irradiance (W/m$^2$) from [9], A is the panel area (m$^2$), and efficiency, $\eta_{PV} = 0.16$ [11].

$$P_t^{PV} = G_t \times A \times \eta_{PV} \tag{7}$$

A 3-by-3 sensitivity study across $E_{max} \, \epsilon\{5.0, 14.4, 28.8\}\, kWh$ and $A \, \epsilon\{16, 28, 40\}\, m^2$ is conducted. Battery sizes were selected to cover the commercially available range of UK residential $LiFePO_4$ systems: entry-level (5 kWh), mid-range (14.4 kWh), and oversized (2× 14.4 kWh) configurations. PV areas represent small, medium, and large UK residential installations.

Following simulation, degradation cost is estimated post-hoc. Rainflow cycle counting [6] is applied to the SOC profile to extract individual cycles, each characterised by its depth of discharge ($DoD_k$). The fraction of battery life consumed per cycle, $f_k$, is defined by equation (8), where $\delta = 4.0253$, $c = 1.0923$ are empirically fitted stress coefficients from $LiFePO_4$/graphite ageing data [4], and $N_{nom} = 5600$ is the nominal cycle life [12]. The total degradation cost over the study period, $C_{deg}$, is given in (9), where $C_{batt} = £500/kWh$ [13]. PV module and inverter losses are not modelled in this study

$$f_k = \frac{\delta \times DOD_k^c}{N_{nom}} \quad (8)$$

$$C_{deg} = \sum_k f_k \times E_{max} \times C_{batt} \quad (9)$$

## 3. RESULTS & DISCUSSION

Over the study period, the household incurs a total import cost of £1,392.74 under the Agile tariff with no PV or BESS installed. Table 1 presents the full 3-by-3 sensitivity study results. Integrating PV and BESS reduces net energy cost across all configurations. Stored energy can be discharged during high-price periods or exported to the grid, enabling price arbitrage under the Agile TOU tariff. In the most favourable configuration (28.8 kWh, 40 m$^2$ PV), the system generates net export revenue of £476.07, compared to the £1,392.74 baseline import cost. However, energy cost alone does not reflect the true cost of system operations.

*Table 1: Energy cost, degradation cost and true total cost across the 3x3 sensitivity study.*

| Configuration | Energy Cost (£) | Deg. Cost (£) | True Cost (£) |
|---|---|---|---|
| No system | 1,392.74 | - | 1,392.74 |
| 5kWh, 16m$^2$ | 601.79 | 60.45 | 662.24 |
| 5kWh, 28m$^2$ | 330.78 | 60.55 | 391.33 |
| 5kWh, 40m$^2$ | 106.19 | 60.65 | 166.84 |
| 14.4kWh, 16m$^2$ | 285.32 | 125.74 | 411.06 |
| 14.4kWh, 28m$^2$ | 13.72 | 125.36 | 139.08 |
| 14.4kWh, 40m$^2$ | -230.94 | 125.99 | -104.95 |
| 28.8kWh, 16m$^2$ | 19.11 | 202.69 | 221.80 |
| 28.8kWh, 28m$^2$ | -239.09 | 197.58 | -41.51 |
| 28.8kWh, 40m$^2$ | -476.07 | 200.09 | -275.98 |

The post-hoc degradation analysis reveals a high hidden cost that the MILP objective does not account for. A key observation is that degradation cost remains constant with battery size, regardless of PV array size. This is because the MPC dispatch strategy cycles the battery similarly as it optimises energy cost only without any degradation penalty. This behaviour is most impactful in configurations where energy import cost is low due to high PV generation or export revenue. In the 14.4 kWh 28 m$^2$ configuration, the energy cost is only £13.72, but the hidden degradation cost is £125.36, accounting for 914% of the energy cost. This behaviour is similar for the 28.8 kWh 16 m$^2$ configuration. In both cases, a naïve energy-cost-only controller would report near grid neutrality, whereas the true system cost substantially exceeds this. Three configurations achieve a negative true total cost: 14.4 kWh/40 m$^2$, 28.8 kWh/28 m$^2$, and 28.8 kWh/40 m$^2$, where export revenue is sufficient to offset both energy and degradation costs.

These results demonstrate that energy-only optimisation systematically underestimates true system costs by ignoring degradation, which becomes the dominant expense in specific configurations. Driven by export-linked arbitrage under the Agile TOU tariff, aggressive cycling can result in degradation cost of up to 1,060% of energy cost, exposing the financial risks of treating batteries as "costless" buffers. The "invisibility" of these costs within the current MILP objective directly motivates the transition toward a more robust, degradation-aware control formulation.

## 4. Conclusion

This study quantifies the hidden degradation cost imposed by energy-cost-only MILP optimisation in a UK residential HEMS using real household demand data and time-of-use tariffs. The results demonstrate that degradation cost is driven primarily by battery size, rather than PV configuration, remains approximately constant regardless of PV array size. Degradation becomes the dominant term in configurations where high PV penetration drives energy cost toward zero and reaches up to 1,060 % of the apparent energy cost in the 28.8 kWh 16 $m^2$ scenario.

A key limitation of the MILP formulation is that perfect foresight is assumed over the 24-hour receding horizon, a condition that does not hold true in practice due to uncertainty in demand, irradiance and tariff prices. Furthermore, the computational cost of MILP scales poorly when a degradation penalty is introduced into the objective function, as the resulting non-linear cost terms cannot be directly handled by a linear solver without approximation.

From a system sizing perspective, these results highlight the importance of matching BESS capacity to both consumption and generation profiles. An undersized battery relative to PV generation undergoes more aggressive cycling, accelerating degradation, whereas an oversized battery relative to available generation incurs capital and degradation costs without proportionate energy benefit. Formal co-optimisation of BESS and PV capacity to balance degradation against energy and export revenue requires further investigation.

These findings reveal that standard HEMS controllers view batteries as costless buffers, neglecting the expense of cycling. Unlike MILP, a reinforcement learning agent can potentially navigate the complex, non-linear relationship between SOC, DoD, and cost without requiring perfect foresight or a convex objective by learning degradation-aware behaviour directly from interaction with the environment.